\documentstyle[aps]{revtex}
\newcommand{\vcc}[1]{\stackrel{\to}{{\bf#1}}}
\newcommand{\vccd}[1]{\stackrel{\to}{{\rm#1}}}

\begin{document}
\title{ Controlling of exciton condensate by external fields and phonon laser}
\author{Yu.E.Lozovik\cite{email},
I.V.Ovchinnikov}
\address{Institute of Spectroscopy, 142090 Moscow reg., Troitsk, Russia}
\maketitle
\sloppy
\begin{abstract}
  The novel method of observation and controlling of Bose-Einstein
condensation in the system of spatially and momentum-space indirect
excitons in coupled quantum wells using in-plane magnetic and
normal electric fields is proposed. Fields are used for
exciton dispersion engeneering. In the presence of in-plane
magnetic field ground state of spatially indirect exciton becomes
also indirect in the momentum space. Manipulation of electric field
magnitude is used for tuning to resonance of transition of
"dark" long-life spatially and momentum-space {\it indirect}
exciton in condensate into spatially and momentum-space {\it
direct} exciton in radiative zone, with phonon creation,
and with subsequent recombination of the direct exciton in
radiative zone. Recombination rate of indirect excitons in
condensate according to proposed scheme sharply rises at resonance.
As a result besides enhanced spectral narrow photoluminescence
connected with recombination of direct exciton in radiative zone
with specific angular dependence, almost monochromatic and
unidirectional beam of phonons appears. An opportunity of
obtaining of "phonon laser" radiation on the basis of this effect
in considered.
\end{abstract}
  The system perspective for observation of different phases of excitons
is the system of spatially indirect excitons (SIE) in coupled
quantum wells (CQW) (see Ref.\cite{1,11,111} and references
therein). The point is that for the system of SIE in CQW - with
spatially separated electron and hole (e and h) - recombination is
suppressed by exponentialy small overlaping of e and h wave
functions. Thus, the case can be easily achieved when the
relaxation times of electrons, holes or excitons is sufficiently
smaller than their lifetime. Therefore, there is an opportunity for
reliable observation of different equilibrium exciton phases. In
superfluid phase of the system of spatially separated e and h
interesting phenomena can be observed, such as nondissipative
electric currents in each quantum well, Josephson effects {\it
etc}. \cite{11,1,2}. Very interesting recent experiments \cite{31,3}
(see also essential previous works \cite{4}) bear witness to an
existence of coherent effects in SIE system at low temteratures.

In this paper we propose new method of the observation and
controlling of coherent phase of SIE by external fields. We
consider direct-gap $^1S$-excitons in $GaAs/AlGaAs$ CQW. In-plane
magnetic field shifts SIE dispersion minimum, where Bose-Einstein
condensate (BEC) of SIE forms, out of the radiative zone. Energy of
SIE in BEC is linear dependent on magnitude of normal electric
field. Thus changing the electric field value enables one to tune
the system of {\it indirect} excitons in condensate to resonance of
process of transformation into spatially and momentum space {\it
direct} excitons (SDE) in radiative zone, by emitting phonons, and
subsequent recombination of the SDE. It occurs that
photoluminescence (PL) of indirect excitons in condensate according
to proposed scheme sharply rises and has specific angular
dependence at resonance. Besides, almost monochromatic and
unidirectional beam of phonons appears. Possibility of phonon laser
creation using this effect is discussed.

  There are four different bound states for a pair of e and h in CQW
in the absence of external fields (Fig.1 a). Two of them, with e
and h in the same quantum well, correspond to SDE placed in one of
two quantum wells. The other two bound states correspond to SIE
with e and h in different quantum wells. As a result of spatial
separation of e and h SIE has normal to CQW electric dipole
momentum $\pm eD$, where $D$ is interwell distance (signs
correspond to two possible locations of e and h in two wells). In
the case of identical quantum wells SDE and SIE have two-fold
degeneracy. We suppose also that the uncertainty of exciton wave
vector due to CQW roughness {\it etc.} is much smaller than all
momenta involved in the problem. In the absense of external fields
SDE is the lowest excitonic state.

  Electric and magnetic fields change the dispersion of SIE
\cite{11,5,6} (Fig.1 b). Normal electric field splits SIE
sublevels as $\Delta\omega=eDE/\hbar$ due to electric dipole -
electric field interaction. It can be shown that moving SIE has
in-plane magnetic dipole momentum $\pm\frac{eDp}{cM}$ normal to its
velocity, where $p$ is in-plane momentum and $M$ is a mass of the
SIE. Thus in parallel magnetic field SIE sublevel dispersion curves
move oppositely apart from the center of Brillouin zone by the
quantity of $K_m=\frac{eDH}{c\hbar}$ due to velocity-selective
magnetic dipole - magnetic field interaction \cite{7}. This was
experimentally observed in the work \cite{6}. SDE level, as in
electric field, remains unchanged and degenerated.

Thus, by in-plane magnetic and normal electric fields one can
change SIE sublevel position in dispersion space. In particular, applying
normal electric field makes it possible for one of the SIE
sublevels to become the ground state of the excitonic system
(instead of SDE at E=0).

  The process of recombination of excitons with emitting a photon
is possible only inside the radiative zone - a small area in the
center of Brillouin zone with $k<k_r\cong E_g/(c\hbar)$, where $E_g$ is
the energy gap in semiconductor, $c$ is the speed of light in the media
and $k_r$ is the boundary wave vector of the radiative zone (in $GaAs$
$k_r\cong 3\cdot10^{5} cm^{-1}$). This is due to photon inability to
carry away an in-plane momentum greater than $k_r$. Out of the radiative
zone the process of photon recombination of the exciton is
forbidden, and the process of exciton recombination with
production of an acoustic phonon and a photon becomes the most
probable exciton recombination process \cite{8}. This
process is of the second order of magnitude of the exciton interaction
with photon and acoustic phonon fields. Greatest parts of energy and
2D momentum of exciton are carried away by photon and phonon,
correspondingly.

  From preceding discussion, the following scheme of the observation of BEC
of SIE can be developed. In the presence of such normal electric
and in-plane magnetic fields that SIE is the ground excitonic state with
minimum of its dispersion curve situated out of the radiative zone,
SIE ground level is being pumped by laser radiation into the radiative
zone (Fig. 2a). After laser switch off the system is evolving using
two channels - one part of pumped excitons in radiative zone
recombines with production of photons, and the other part of excitons
relaxate into the minimum of dispersion of SIE level out of the
radiative zone, tranferring their extra energy and momentum to
acoustic phonons (Fig. 2b). Thereafter, the system of SIE reaches a
quasi-equilibrium state with the temperature of reservoir (lattice),
and in case of sufficiently low temperature of lattice the system of SIE forms
BEC. We will consider $T=0$ for simplicity; at
temperatures much less than Kosterlitz-Thouless temperature optical
properties are changed only slightly with respect to those at $T=0$.
The state of the system at this moment of evolution is the state
where almost all excitons are at SIE level in its dispersion minimum with wave
vector $\vccd{K}_m$ (for low density of excitons) and all the other
levels are not populated. In high
magnetic field we have $K_m\gg k_r$ ({\it e.g.} $H=10T$ and $D=10nm$ corresponds to
$K_m=2\cdot10^6 cm^{-1}$).

Bose-condensed SIE situated in momentum space at $K=K_m$ position are "dark",
that is the process of emitting of photons is weak. The rate of
recombination SIE in BEC through intermediate production of virtual exciton in
radiative zone (with production of photon and phonon with wave vectors
in the region of momentum space $d^3 q, d^3 k$) is
{\footnotesize
\begin{eqnarray}
&dW_{phn,pht}=
\left(\sum\limits_s\frac{(M_{s,pht}M_{s,phn})^2}{(\omega_{pht}(\vcc{k})-
\omega_{s}(\vccd{k}))^2+\eta_s^2}+ \sum\limits_{s_1,s_2,s_1\ne s_2}
\left(\frac{M_{s_1,pht}M_{s_1,phn}M_{s_2,pht}M_{s_2,phn}}
{2(\omega_{pht}(\vcc{k})-\omega_{s_1}(\vccd{k})+i\eta_{s_1})
(\omega_{pht}(\vcc{k})-\omega_{s_2}(\vccd{k})-i\eta_{s_2})}+c.c.\right)
\right)\nonumber\\
&\delta(\vccd{k}+\vccd{q}-\vccd{K}_m)
\delta(\omega_{pht}(\vcc{k})+\omega_{phn}(\vcc{q})-\omega_{{SIE}}(\vccd{K}_m))
(2\pi)^{-3}\rho N_{\vcc{k}},N_{\vcc{q}} d^3 q d^3 k \label{Wphtphn}
\end{eqnarray}
} In Eq.(\ref{Wphtphn}) $\rho$ is 2D density of SIE in CQW;
$N_{\vcc{k}},N_{\vcc{q}}$ are numbers of photons and phonons in the
states with wave vectors $\vcc{k}$ and $\vcc{q}$, correspondingly;
notation $\vccd{x}$ implies 2D in-plane vector $x$ or in-plane
component of 3D vector $\vcc{x}$; $M_{s,phn}$ is the matrix element
of transformation of SIE into $s$-state exciton in radiative zone
with acoustic phonon creation, and $M_{s,pht}$ is the matrix
element of photon recombination of $s$-exciton in radiative zone;
$\eta_s$ is the width of $s$th exciton level. Using $GaAs$
parameters, we estimate for SDE $\eta_{SDE}\cong 10^{10} sec^{-1}$.
Recombination rate consists of resonant terms, corresponding to the
processes with intermediate creation of virtual s-exciton in
radiative zone, and nonresonant terms.

  Main contribution to the rate of SIE recombination process is
originated from recombination transitions, in which virtual
s-excitons are mostly close to their mass shell. SIE and SDE are
such s-excitonic states in our case. At this moment of evolution
(Fig.2b) the process goes mainly through virtual SIE in radiative
zone. For the estimation we take phonon energy to be equal to
$\omega_{phn}(\vcc{q})\approx c_vK_m$, where $c_v$ is a speed of
sound. In this case $\omega_{pht}(\vcc{k})\approx
\omega_{{SIE}}(\vccd{K}_m) - c_vK_m$ and we get: {\footnotesize
\begin{eqnarray}\label{Wsim}
W_{phn,pht}\sim
\frac{(M_{SDE,pht}M_{SDE,phn})^2}
{(c_vK_m - (\omega_{{SIE}}(\vccd{K}_m)-\omega_{{SDE}}(0)))^2+\eta_{SDE}^2}+
\frac{(M_{SIE,pht}M_{SIE,phn})^2}
{(c_vK_m - (\omega_{{SIE}}(\vccd{K}_m)-\omega_{{SIE}}(0)))^2+\eta_{SIE}^2}
\end{eqnarray}
}
  In this formula we also neglected non-resonant terms. Parameter
$(\omega_{{SIE}}(\vccd{K}_m)-\omega_{{SIE}}(0))$ is not dependent on
electric field magnitude, while $(\omega_{{SIE}}(\vccd{K}_m)-
\omega_{{SDE}}(0))$ is a linear function of electric field magnitude,
and equals zero when the electric field has a particular (resonant)
value. For CQW system used in Ref.\cite{6} at magnetic field $10T$
these parameters are correspondingly equal to $0.8\cdot 10^{13}
sec^{-1}$ and $3.5 \cdot 10^{13} sec^{-1}$.

  At this stage of the system evolution one can reduce the magnitude of
electric field down to the moment, when the conservation laws for
energy and 2D momentum are satisfied in the processes of SIE in BEC
transformation into real SDE in radiative zone with production of
acoustic phonon and SDE in radiative zone photon recombination (Fig.2c)
(in other words, when resonance condition is satisfied).
Using Eq.(\ref{Wsim}), one can get the estimation for the ratio of
PL rates at initial condition and after tuning to the resonance in
the system used in Ref.\cite{6}:
{\footnotesize
\begin{eqnarray}
\frac{W_{resonance}}{W_{initial}}\approx
\alpha^2\frac{(\omega_{{SIE}}(\vccd{K}_m)-\omega_{{SIE}}(0))^2}
{\eta_{SDE}^2} \approx 10^{5 \div 6}\alpha^2,\hspace{.2cm}
\alpha=\frac{M_{SDE,pht}}{M_{SIE,pht}}
\frac{M_{SDE,phn}}{M_{SIE,phn}}
\label{enhancement}
\end{eqnarray}
}
  Martix elements of phonon-exciton interaction vertex are determined
by Bardeen-Shockley deformational potential Ref.\cite{10}.
Matrix element of photon recombination of an exciton in radiative zone
is proportional to the overlaping integral of e and h in the exciton
\cite{8}. Using Eq.(\ref{enhancement}) one can show that
PL intensity can be increased, by changing
the electric field, at least by two orders. Therefore, after tunning
the system to the resonance the recombination rate will be greatly
increased, and this will result in sharp PL intensity growth.

 Now we will analize angular dependence of resonant PL.
Let magnetic field be directed along x-axis. In this case $\vccd{K}_m$
vector is parallel to y-axis.
Studying resonant radiation, in Eq.(\ref{Wphtphn}) we take only
resonant term responsible for SIE recombination by intermediate SDE
creation. Integrating (\ref{Wphtphn}) over $d^3 q$, we obtain
the rate of photon emission in k-space area $d^3 k$:
{\footnotesize
\begin{eqnarray}
dW_{pht}^{res}&\sim&
\frac{1}{\left[(\omega_{pht}(\vcc{k})-
\omega_{{SDE}}(\vccd{k}))^2+\eta_{SDE}^2\right]} \frac{(\omega_{{SIE}}(\vccd{K}_m)-
\omega_{pht}(\vcc{k}))d^3 k} {\sqrt{(\omega_{{SIE}}(\vccd{K}_m)-
\omega_{pht}(\vcc{k}))^2-c_v^2(\vccd{K}_m-\vccd{k})^2}} \label{Wpht}
\end{eqnarray}
}
  The maximum energy of the photon is $\hbar
c\frac{\omega_{{SIE}}(\vccd{K}_m) - c_v K_m}{c+c_v}$. Photons with
any energy less than its maximum value can be created in the
process, but the rate of production of photons with small energy is
suppressed by Lorentz factor.

  The process becomes resonant in the case when the maximum of the
Lorentz factor and the singularity of the second factor coincide
for at least some values of $\vccd{k}$. This condition can be made
clear graphically (Fig.3a). Resonant condition is satisfied if in
the space ($\vccd{p},\omega$) a dispersion cone of 2D phonon, drawn
down from BEC position, intersects SDE dispersion surface in the
radiative zone. We stress, that 2D dispersion of 3D phonons gives
only the minimum energy of a phonon with a given component of 3D
wave vector on CQW plane (the same is true for photons). The
minimum energy corresponds to zero normal to CQW component of
phonon wave vector. For this reason, resonant process with SIE
recombination with intermediate creation of real SDE takes place
even in case of greater energy of exciton in BEC than resonance one
({\it i.e.} in smaller elecric fields), but with less intensity.

  The resonance condition corresponds to a distinct range of
electric field magnitude. We represent exciton in BEC energy dependence
on electric field magnitude as $\omega_{{SIE}}(K_m)=c_v(K_m+1/2 k_r
T(E))$, where $T(E)$ is a linear function of electric field. Since $K_m\gg
k_r$, we can admit that in the case of resonance, 2D phonon dispersion
(cone) intersects the SDE dispersion surface in a circle (Fig.2 b):
   \begin{eqnarray}
(k_x-k_0)^2+k_y^2=k_rk_0T(E)+k_0^2,\hspace{.2cm}k_0=\frac{Mc_v}{\hbar}\label{krug}
   \end{eqnarray}
  For $GaAs$ we have $k_0=10^5cm^{-1}$. The condition of the resonance
can be represented as $2+\frac{k_r}{k_0}\ge T(E)\ge -\frac{k_0}{k_r}$.

  We will below admit resonance approximation, {\it i.e.} substitution for
the Lorentz factor by delta-function $\delta(\omega_{pht}(\vcc{k})
-\omega_{{SDE}}(\vccd{k}))$. This approximation is valid when
$\eta_{SDE}\ll c_v k_r$, so that it is applicable in our case since
$c_vk_r=1.5\cdot 10^{11}sec^{-1}$. In result one can get:

\begin{eqnarray}
dW_{pht}^{res}(\theta,\phi) \sim \frac{d\Omega} {\sqrt{T(E)-\beta
sin^2(\phi)sin^2(\theta)-cos(\theta)}} ,\hspace{0.2cm}
\beta=k_r/k_0\cong 3 \label{uglzav}
\end{eqnarray}

  Here $d\Omega$ is the spatial angle, $\theta$ and $\phi$ are azimuth
and polar anlges of spherical coordinates in photon wave vector
$\vcc{k}$ space. Azimuth axis ($\theta=0$) is directed parallel to
$\vccd{K}_m$. Eq. (\ref{uglzav}) gives a resonant radiation angular
dependence. Resonant PL is absent when $T(E)<-k_0/k_r$. At
resonance ($2+k_r/k_0\ge T(E)\ge -k_0/k_r$) resonant photons are
emitted into spatial angle, which is formed by intersection of
sphere with radius $k_r$ and a cylinder with base (\ref{krug}) and
generatrix parallel to z-axis (Fig.3 c). In the resonant
approximation on boundary of this spatial angle the rate of
resonant PL has an integrating singularity. It can be shown that
real rate of PL right on this boudary is approximately as
much as $\frac{c_vk_r}{\eta_{SDE}} \approx 15$ times greater than
in the other area of spatial angle of the resonant radiation. If
$T(E)>2+k_r/k_0$ resonant PL intensity has no singularity on the
boundary of spatial angle of resonant PL, and with increasing
$T(E)$ (reducing electric field) tends to become more and
more isotropic.

  When resonance condition is satisfied, energy level scheme of our
system is similar to that of three-level {\it impulse} one-pass
laser with inverse population of upper level and with rapid lower
transition (Fig.4 a). Rapidity of the lower transition with respect
to the the rate of upper transition is supplied by tunneling
character of transformation of SIE into SDE with creation of
acoustic phonon. The difference is that in our case phonons are
emitted instead of photons. Wave vectors of resonant phonons form a
prolate ellipsoid of revolution with base (\ref{krug}) and the
ratio of radii $\sqrt{K_m/k_0}$ (Fig.4 b). The best quality of
unidirectivity and monochromatism of the resonant phonons is
evidently achieved in electric field that corresponds to the
relation $T(E)=-k_0/k_r$, when the ellipsiod reduces to a point, so
that for discussion of phonon "laser" radiation we will consider this
case.

At this resonance condition the coherence of resonant phonons at
early stages of the process of phonon emission is determined by
total (in a sense "homogeneous" and "inhomogeneous") width of
phonons. Width of wave vector distribution of phonons can be
represented as $\Delta q=\sqrt{k_0(\Delta q_{hom} + \Delta
q_{inh})}$, where $\Delta q_{hom}$ is due to widths of SDE level
and phonon state, and $\Delta q_{inh}$ is connected with SIE
momenta spread that corresponds to depleting of BEC due to
interactions of excitons; $\Delta q_{inh}=\sqrt{U(0)\rho M}/\hbar$,
where $U(0)$ is zero Fourier component of interaction between SIE;
$\Delta q_{hom}=(\eta_{SDE}+\eta_{phn})c_v^{-1}$, where
$\eta_{phn}$ is the phonon attenuation rate.

There are two sufficient conditions for phonon radiation to be the
"laser" one: generation condition and condition of macroscopic
population of each quantum state. In our case the first condition
is the prevailing rate $\eta_{spon}$ of spontaneous phonon emission
in the process over the rate of phonon attenuation in the sample
$\eta_{phn}$. Attenuation of the phonon is determined by the media
properties, temperature, the size of the sample and scattering of
phonon on the planes of CQW. Using data from Ref.\cite{6} and
Eq.(\ref{enhancement}) one can get, that $\eta_{spon}$ can be not
simply greater than the phonon attenuation rate, but even can reach
the phonon energy $K_m c_v\cong 10^{12}sec^{-1}$.

The second condition is macroscopic population of each phonon
quantum state (when all SIE in BEC are recombined and thus have
contributed to phonon radiation). Phonon radiation occupies
$V(\Delta q)^3/(2\pi)^3$ quantum states, where $V$ is a volume of
the sample, while the number of SIE is $S\rho$, $S$ is the area of
CQW. The condition is $L_z\ll (2\pi)^3\rho/(k_0(\sqrt{U(0)\rho
M}/\hbar+\Delta q_{hom}))^{3/2}$, where $L_z$ is the width of the
sample in $z$-direction. The quantity $L_z$ appeared due to that
phonons and SIE differ in their dimensions (3D phonons {\it vs.} 2D
excitons).

In conclusion we emphasize that discussions above point out
possible existence of interesting effect - "phonon laser" (details
will be published elsewhere). Using all advantages of the analogy
with photon laser, one can propose to adjust a phonon resonator for
the purpose of increasing phonon coherence. The resonator can
experimentally be realized as phonon mirrors (the media with
greater speed of sound) on (y-direction) edges of the sample. In
this case one-pass phonon "laser" becomes multi-pass one. The
restriction for the width of the sample (second condition) can be
made weaker by considering heterostructure consisting of many CQW
(the distance between CQW must satisfy the above condition for
$L_z$).

Note that, coherent phonons modulate the dielectric function of the
media. This gives an opportunity of detection of the effect of
coherent phonon generation by study the modulation of optical
properties of the media, which can be observed by time-resolved
femtosecond spectroscopy ({\it e.g.} by pump-probe method). Another
possibility to observe coherent properties of phonon radiation
is analysis of its statistics, {\it e.g.} by Hunbury-Brown and Twiss
method.

  The work has been supported by Russian Foundation of Basic
Research, INTAS and Programm "Solid State Nanostructures".

\appendix

\newpage
Captures to Figures.\\
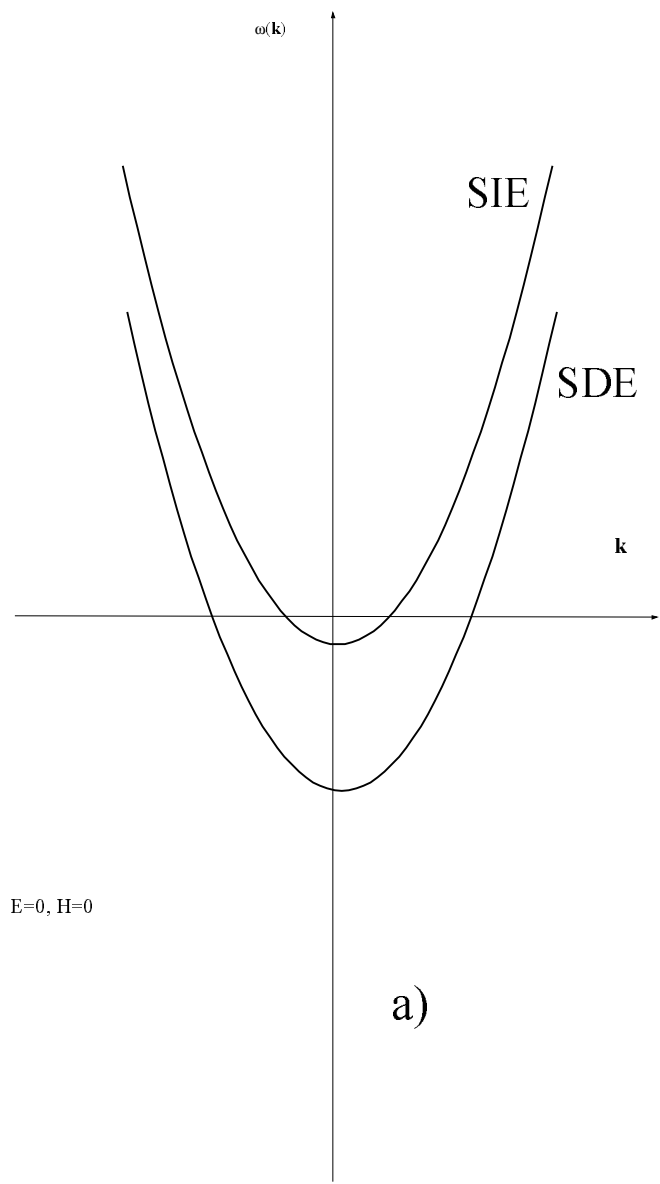
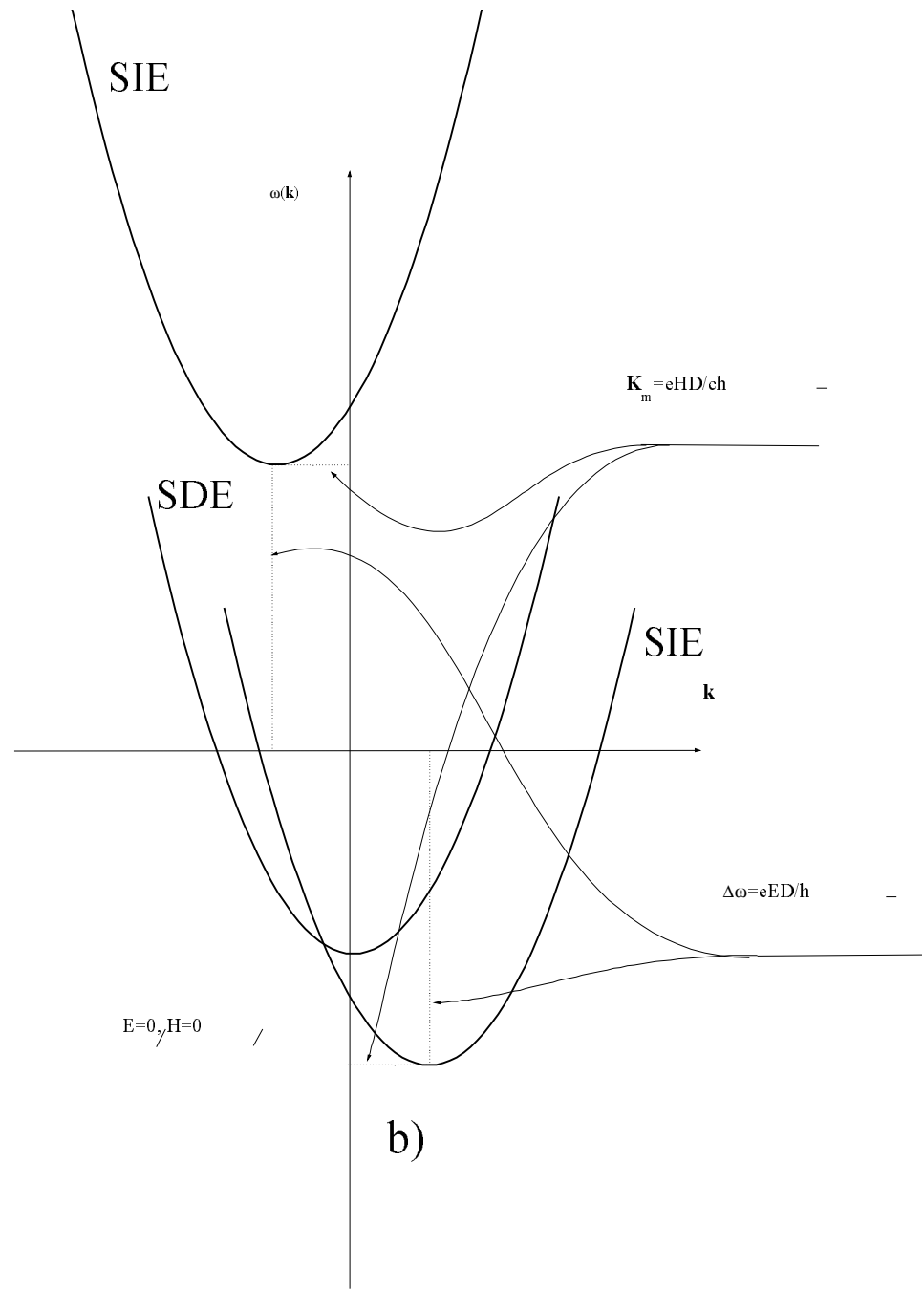
{\it Fig.1}\hspace{0.5cm}
Dispersion laws of direct and indirect excitons in coupled quantum wells
a) Without extrenal electromagnetic fields SDE level is lower than SIE
level by the difference of Coloumb interactions in SDE and SIE \hspace{0.3cm}
b) In parallel magnetic field SIE sublevel dispersions move oppositely
apart by wave vector $K_m=\frac{eHD}{c\hbar}$; in normal electric
field one of SIE sublevels lowers as $\Delta \omega = eED/\hbar$
and the other rises as $\Delta \omega$\\
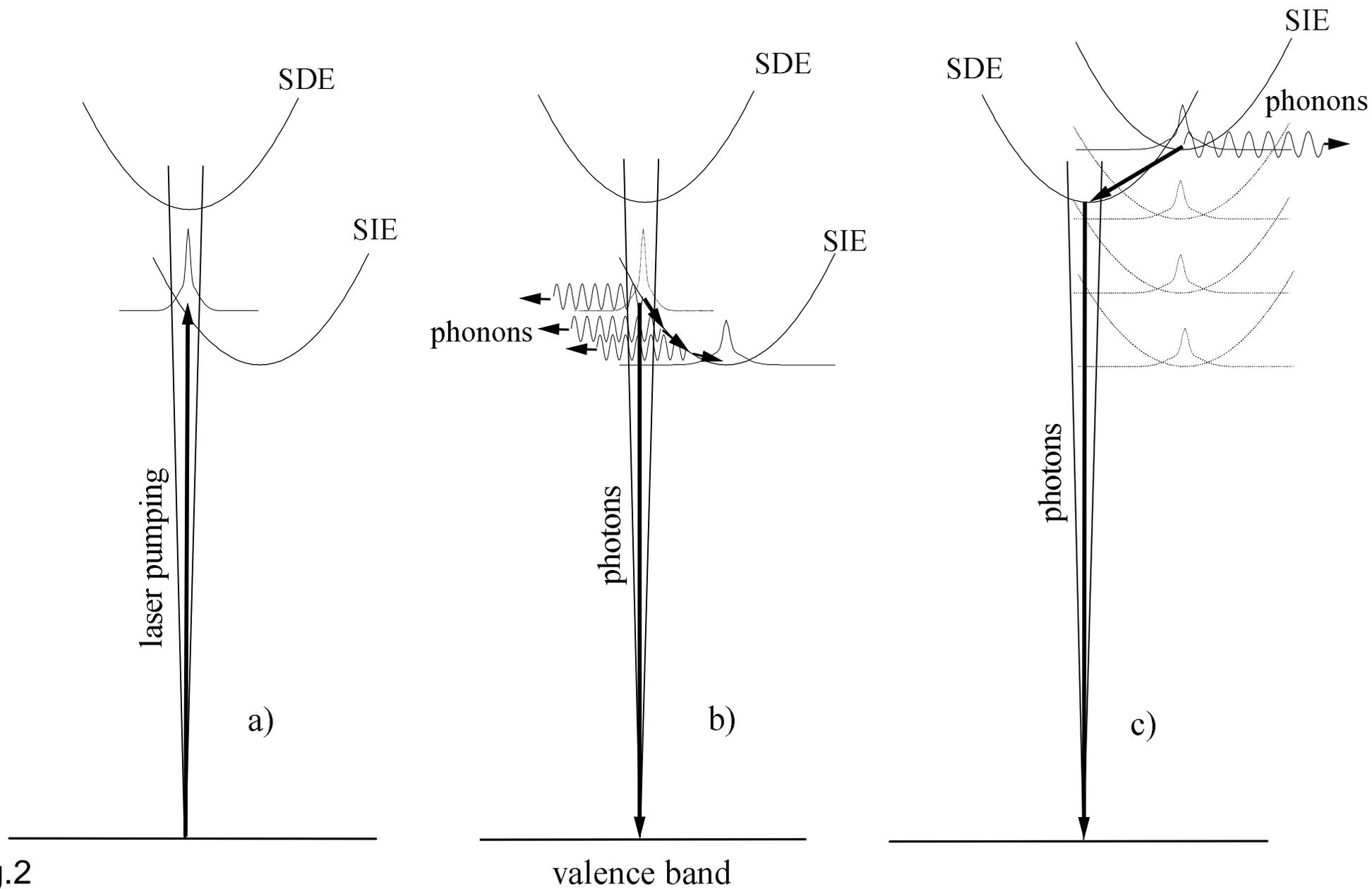
{\it Fig.2} \hspace{0.5cm}
a) Laser pumping of SIE level \hspace{0.3cm} in the
presence of parallel magnetic and normal electric fields.
$K_m=\frac{eHD}{c\hbar}$ is the displacement of dispersion minimum in
parallel magnetic field, $k_r$ is boundary wave vector of radiative zone.
b) Relaxation of indirect excitons. One part of excitons recombines
with production of photons, and the other part relaxates to the minimum
of SIE dispersion $K_m$.\hspace{0.3cm}
c) Changing of BEC position when electric field magnitude is being reduced.\\
{\it Fig.3} \hspace{0.5cm}
a) Position of direct and indirect exciton dispersions when condition of
the resonance is satisfied.\hspace{0.3cm}
b) intersection of dispersion surface of indirect excitons with 2D phonon
dispersion.\hspace{0.3cm}
c) segments of spatial angle correspond to the boundary of resonant
photoluminescence angular dependence.\\
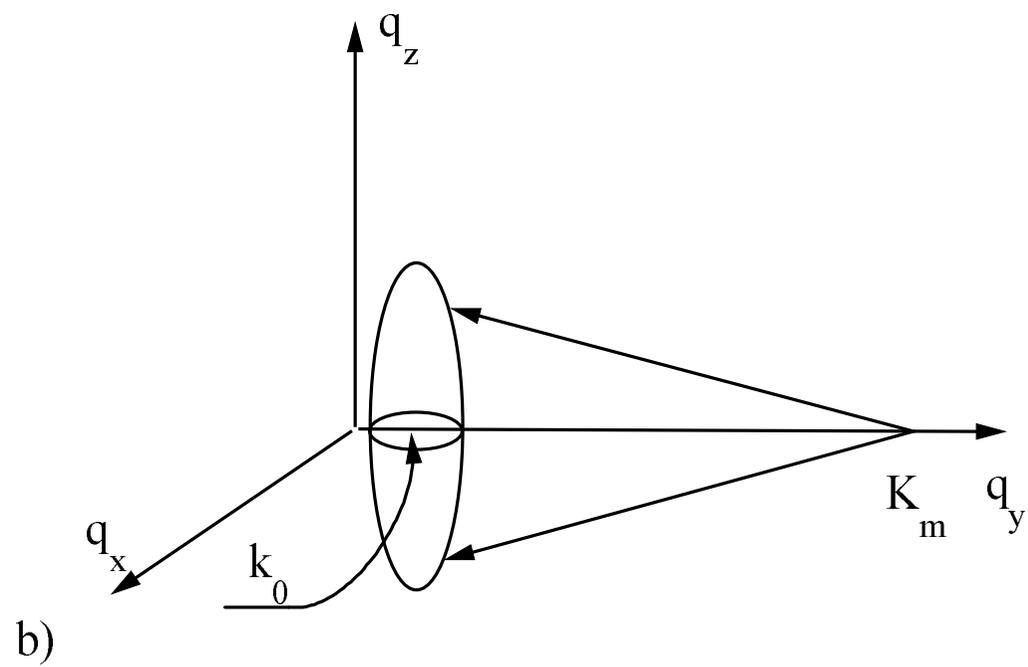
{\it Fig.4} \hspace{0.5cm} a) Three-level laser analog of phonon laser.
\hspace{0.3cm} b) Phonon wave vector distribution. Wave vectors of
phonons form prolate ellipsond of revolution with ratio of its
radii $\sqrt{\frac{K_m}{k_0}}$.

\begin{thebibliography}{99}
\bibitem[*]{email} e-mail: lozovik@isan.troitsk.ru, lozovik@mail.ru
\bibitem{1} Yu.E.Lozovik, O.L.Berman, JETP Lett., {\bf 64}, 526 (1996);
Yu.E.Lozovik, O.L.Berman, V.G.Tsvetus, Phys.Rev.B, {\bf 56}, 5628
(1999).
\bibitem{11}
Yu.E.Lozovik, V.I.Yudson, JETP Lett. {\bf 22}, 36 (1975);
Sol.St.Comms., {\bf 18}, 628 (1976);
Xu.Zhu, P.B.Littlewood, M.S.Hybertsen, and T.M.Rice, Phys.Rev.Lett., {\bf74}, 1633 (1995);
Y.Naveh, B. Laikhtman, Phys.Rev.Lett., {\bf 77}, 900 (1996);
A.Imamo\=glu, Phys.Rev.B, {\bf 57}, R4195 (1998);
S.Conti, G.Vignale, A.H.MacDonald, Phys.Rev.B, {\bf 57}, R6846 (1998);
I.V.Lerner, Yu.E.Lozovik, JETP, {\bf 80}, 1488 (1981) [Sov.Phys. JETP, {\bf 53}, 763 (1981)];
D.Yoshioka, A.H.MacDonald, J. Phys. Soc. Jpn., {\bf 59} 4211 (1990);
J.B.Stark, W.H.Knox, D.S. Chemla et.al., Phys.Rev.Lett., {\bf 65}, 3033 (1990);
X.M.Chen,J.J.Quinn, Phys.Rev.Lett. {\bf 67}, 895 (1991);
G.E.W.Bauer, Phys.Scripta, T{\bf 45}, 154 (1992).
\bibitem{111} See also excitonic phases in 3D:
L.V.Keldysh, Yu.V.Kopaev, Fiz. Tverd. Tela, {\bf 6}, 2791 (1964)
[Sov.Phys.Solid State, {\bf 6},2219 (1965)];
B.I.Halperin, T.M.Rice, Solid State Phys., {\bf 21}, 115 (1968);
"Bose Einstein Condensation", eds. A.Griffin, D.W.Snoke,
S.Stringrari, Cambridge Univ. Press, Cambridge (1995);
A.L.Ivanov, H.Haug, L.V.Keldysh, Phys.Rep., {\bf 296}, 237, (1998).
\bibitem{2} A.V.Klyuchnik,Yu.E.Lozovik, J.Low Temp. Phys., {\bf 38}, 761 (1980);
J.Phys.C {\bf 11}, L483 (1978);
S.I.Shevchenko, Phys.Rev.Lett., {\bf 72}, 3242 (1994);
Yu.E.Lozovik, A.V.Poushnov, Phys.Lett.A, {\bf 228}, 399 (1997).
\bibitem{31} L.V.Butov, A.I.Filin, Phys.Rev.B, {\bf 58}, 1980 (1998).
\bibitem{3} A.V.Larionov, V.B.Timofeev, J.Hvam, C.Soerensen, JETP Lett., {\bf 71}, 174 (2000).
\bibitem{4} T.Fukuzawa, E.E.Mendez, J.M.Hong, Phys.Rev.B, {\bf 64}, 3066 (1990);
V.Sivan, P.M.Solomon, H.Strikman, Phys.Rev.Lett., {\bf 68}, 1196 (1992);
L.V.Butov, A.Zrener, G.Abstreiter, G.Bohm, and G.Weimann, Phys.Rev.Lett., {\bf 73}, 304 (1994);
J.-P.Cheng, J.Kono, B.D.McCombe et.al., Phys.Rev.Lett., {\bf 74}, 450 (1995).
\bibitem{5} A.A. Gorbatsevich, I.V. Tokatly, Semicond.Sci.Technol., {\bf 13}, 288 (1998).
\bibitem{6} L.V.Butov, A.V. Mintsev, Yu.E. Lozovik, K.L. Campman, and A.C. Gossard, Phys.Rev.B.,
(in print).
\bibitem{7} This effect can be also treated as diamagnetism of spatially separated
e-h system in CQW \cite{11}.
\bibitem{8} J.Feldmann, G.Peter, E.O.Gobel, P.Dawson, K.Moore, C.Foxon, R.J.Elliot, Phys.Rev.Lett.,
{\bf 59}, 2337(1987);
L.Schultheis, A.Honold, J.Kuhl,~and~K.K\"ohler, Phys.Rev.B, {\bf 33}, 8 (1986).
\bibitem{10} J.Bardeen, W.Shockley, Phys.Rev., {\bf 80}, 72 (1950); A.Anselm, Yu.A.Firsov,
JETP, {\bf 28}, 151 (1955).
\end{thebibliography}
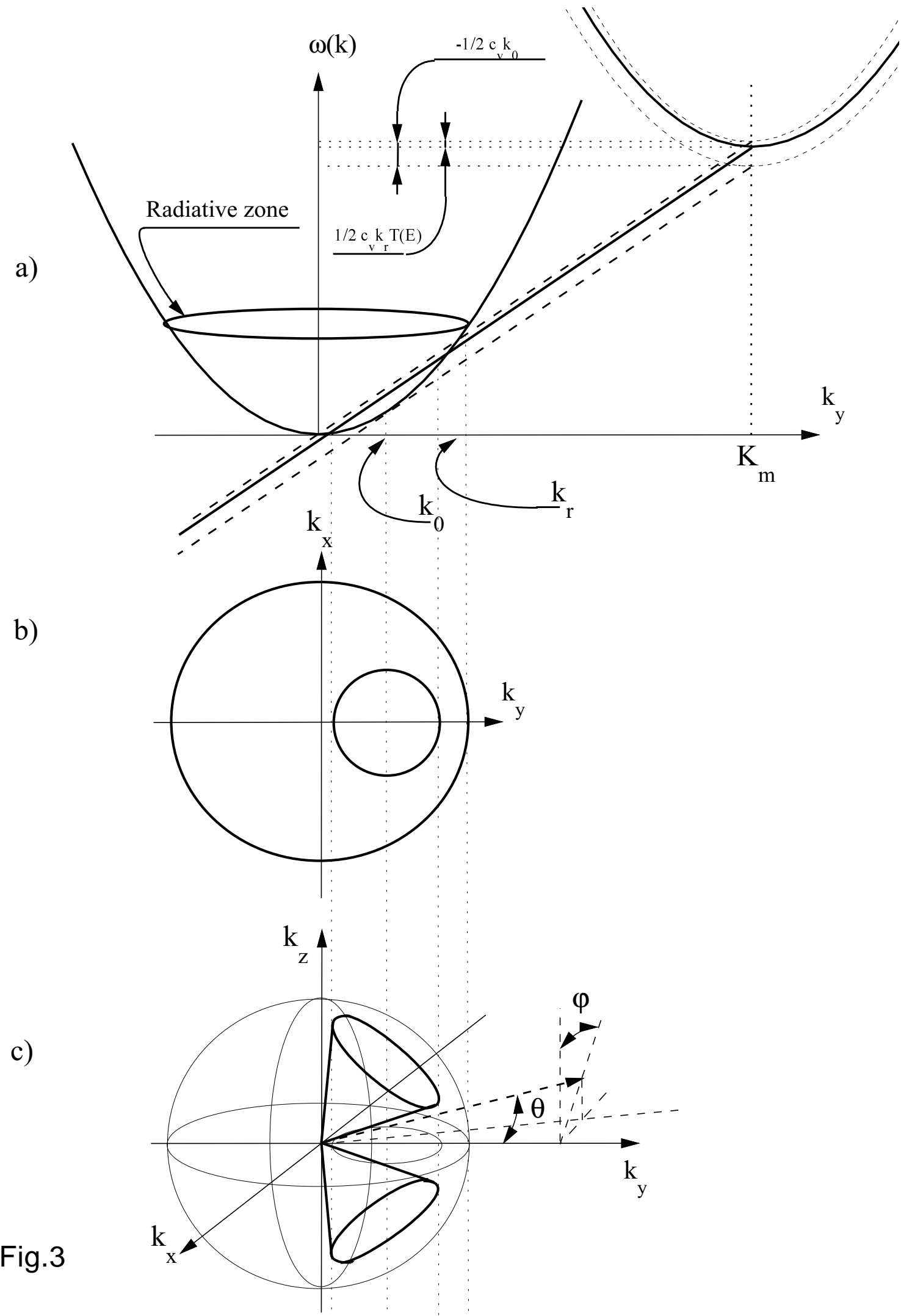
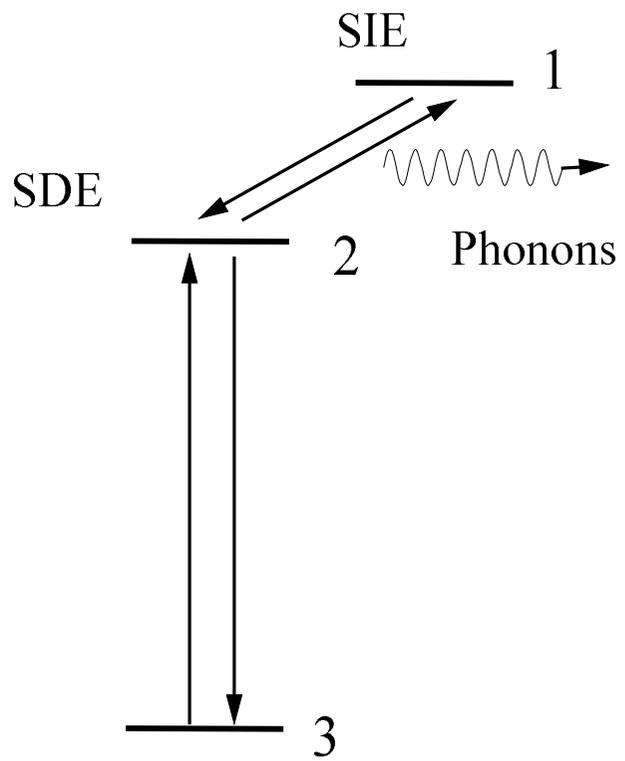
\end{document}